\documentclass[twocolumn, tighten, times]{aastex63}
\usepackage{epsfig}			
\usepackage{graphicx,color}		
\usepackage{amssymb}			
\usepackage{color}			
\usepackage{url}			
\usepackage{amsmath}			
\usepackage{rotating}			
\usepackage{float}			
\usepackage{textcomp}
\usepackage{float}
\usepackage{epstopdf}
\usepackage{dcolumn}
\usepackage{times}
\usepackage{tabularx}
\usepackage[english]{babel}
\usepackage{gensymb}
\usepackage{hyperref}
\usepackage{cleveref}
\usepackage[normalem]{ulem}
\graphicspath{{fig/}}

\newcommand{\Fig}[1]{Figure~\ref{#1}}
\newcommand{\Eq}[1]{Equation~(\ref{#1})}
\newcommand{\angles}[1]{\ensuremath{\left\langle #1 \right\rangle}}
\newcommand{\Bmax}{$B_{\rm max}$}

\def\bl{Babcock--Leighton}

\begin{document}
\title{Magnetic field dependence of bipolar magnetic region tilts on the Sun: Indication of tilt quenching}

\correspondingauthor{Bibhuti Kumar Jha}
\email{bibhuti.kj@iiap.res.in, bibhuraushan1@gmail.com}

\author[0000-0003-3191-4625]{Bibhuti Kumar Jha}
\affiliation{Indian Institute of Astrophysics, Koramangala, Bangalore 560034, India}

\author[0000-0002-8883-3562]{Bidya Binay Karak}
\affiliation{Department of Physics, Indian Institute of Technology (Banaras Hindu University), Varanasi, India}
\affiliation{Indian Institute of Astrophysics, Koramangala, Bangalore 560034, India}

\author[0000-0002-7762-5629]{Sudip Mandal}
\affiliation{Max-Planck-Institut f\"ur Sonnensystemforschung, Justus-von-Liebig-Weg 3, D-37077 G\"ottingen, Germany}

\author[0000-0003-4653-6823]{Dipankar Banerjee}
\affiliation{Indian Institute of Astrophysics, Koramangala, Bangalore 560034, India}
\affiliation{Center of Excellence in Space Sciences India, IISER Kolkata, Mohanpur 741246, West Bengal, India}
\affiliation{Aryabhatta Research Institute of Observational Sciences, Nainital-263002, Uttarakhand, India}

\begin{abstract}
The tilt of bipolar magnetic region (BMR) is crucial in the Babcock--Leighton process for the generation of the poloidal magnetic field in Sun. Based on the thin flux tube model of the BMR formation, the tilt is believed to be caused by the Coriolis force acting on the rising flux tube of the strong toroidal magnetic field from the base of the convection zone (BCZ).
We analyze 
the magnetic field dependence of BMR tilts using the magnetograms 
of \textit{Michelson Doppler Imager} (MDI) (1996--2011) and \textit{Helioseismic and Magnetic Imager} (HMI) 
(2010--2018). We observe that the distribution of the maximum magnetic field ($B_{\rm max}$) of BMRs  
is bimodal. Its first peak at the low field corresponds to BMRs which do not
have sunspots as counterparts in the white light images, whereas the second peak corresponds to sunspots as recorded in both type of images.
We find that the slope of Joy's law ($\gamma_0$) initially increases slowly with the increase of $B_{\rm max}$.
 However, when $B_{\rm max} \gtrsim 2$~kG, $\gamma_0$ decreases.
Scatter of BMR tilt around Joy's law systematically decreases with the increase of $B_{\rm max}$.
The decrease of observed $\gamma_0$ with $B_{\rm max}$ provides a hint to a nonlinear tilt quenching
in the Babcock--Leighton process. 
We finally discuss how our results may be used to make a connection with the thin flux tube model.
\end{abstract}

\section{Introduction}
\label{sec:intro}
Sunspots are the regions of concentrated magnetic field observed as dark spots in white-light images. In the magnetograms, we find two regions of opposite
polarities appearing close to each other. Thus the sunspots that we see in white light image are essentially two poles of a more general feature called the Bipolar Magnetic Regions (BMR). However, the weaker BMRs produce negligible 
intensity contrast and hence go undetected in white light images.
In general, BMRs are tilted with respect to the equator and statistically, this tilt increases with latitude---popularly known as Joy's law \citep{Hale19}. 

The tilt is crucial for the generation of the poloidal magnetic field through the decay and dispersal of the BMRs near the solar surface, 
which is popularly known as the \bl\ process. While this was proposed in the 60s by \citet{Ba61} and \citet{Le64},
%
%
%
in recent years, this process has received significant attention 
due to its support from observational studies \citep{Das10,KO11,Muno13,Priy14}. 
 Based on this \bl\ process, several surface flux transport models have been constructed, 
which are successful in reproducing many features of the solar surface \citep{Ji14}.
Many dynamo models, including the popular flux transport dynamo models, 
have also been constructed based on this \bl\ process 
\citep{Le69,WS91,WSN91}; see reviews \citep{Cha10,Kar14a,C18}. 

A serious concern in these \bl\ models is the saturation of magnetic field. There
must be a nonlinear quenching to suppress the growth of magnetic field in any kinematic dynamo model such as the \bl\ ones.
In the latter models, large-scale velocities, namely, meridional flow and differential rotation are specified (broadly through observations), while the small-scale velocity is parametrized such as in the form of turbulent diffusivity. 
Therefore, the most obvious choice in these models is to include a nonlinearity in the \bl\ process. 
In all the previous \bl\ dynamo models, a magnetic field dependent quenching 
is included such that 
the poloidal field production is reduced when the toroidal magnetic field exceeds the so-called saturation field $B_0$ \citep{Cha10}.
 For the \bl\ process, this requires that the tilt must be reduced when the BMR field strength
exceeds a certain value; see \citet{LC17,KM17,KM18} for specific requirement of this idea. 

\begin{figure*}[!htbp]
    \centering
    \includegraphics[width=0.8\textwidth]{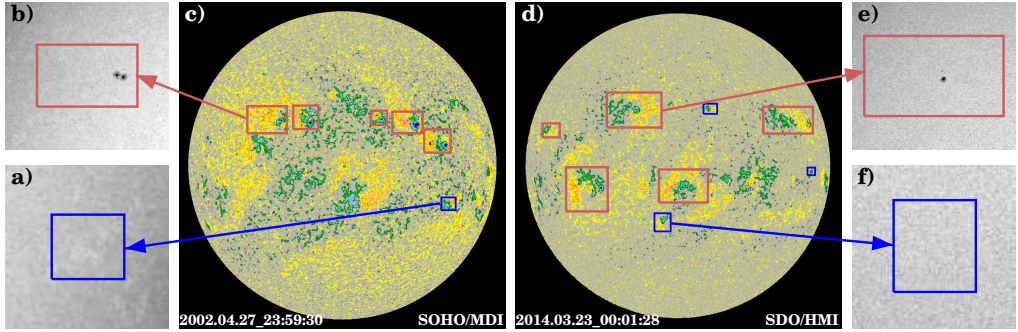}
    \caption{Representatives magnetograms of (c) MDI and (d) HMI (saturated to $\pm$1.5 kG) with BMR$_{\rm{WS}}$ (red box) and BMR$_{\rm{NS}}$ (blue). (a, b, e, and f): show IC counterparts.}
    \label{fig:cont}
\end{figure*}

We believe that the BMRs are produced due to buoyant rise of the strong toroidal magnetic flux tubes from the base of the convection zone (CZ) \citep{Pa55b}.  From the thin flux tube model, we know that during the rise of toroidal flux in the CZ, 
the Coriolis force induced by the diverging east-west velocity near the loop apex causes a tilt \citep{DC93,FFM94}. 
Therefore, we expect the rise time of toroidal flux tube and thus the tilt to decrease with increase of magnetic field in the tube. 
This idea can potentially lead to a quenching in the \bl\ process. 

Although the thin flux tube model explains some observed features of BMRs, it does not capture the detailed dynamics of solar CZ. Indeed,
including the convection, \citet{WFM11} find a significant change in the
behaviour of BMR tilt. They find the tilt to increase with the magnetic field first and then decrease in accordance with the thin flux tube model.

Using magnetogram data corresponding to 1988--2001, 
\citet{Tian03} found a systematic variation of the BMR tilt 
with the magnetic flux content. Surprisingly,
using {\it Michelson Doppler Imager}(MDI) magnetograms during 1996 -- 2011, \citet{SK12} did not
find any systematic variation of the BMR tilt with the magnetic flux
and they claim that their result rules out the thin flux-tube model.
However, we should not forget that the magnetic field of BMR 
also vary with the magnetic flux \citep{Tla14},
and in the analysis of \citet{SK12}, the variation of magnetic field is ignored.
Therefore, the motivation of the present Letter is first to analyse the BMRs 
based on their magnetic field strength. 
Then we shall check how the tilt changes with the magnetic field strength
and whether there is any quenching in the tilt to support the theoretical models of 
BMR formation and the \bl\ dynamo saturation.

\section{Data and Method}
\label{sec:method}
In this work, we have used the full disk Line of Sight (LOS) magnetogram with cadence of 6 hours and Intensity Continuum (IC) with caence of 24 hours from {\it Michelson Doppler Imager} \citep[MDI: 1996--2011;][]{1995SoPh..162..129S} and {\it Helioseismic and Magnetic Imager} \citep[HMI: 2010--2018;][]{2012SoPh..275..229S} for identification of BMRs.

The magnetograms taken from these two instruments, give only the LOS component of magnetic field. To get the magnetic field in the direction normal to the solar surface we have corrected  for the projection effect.
The projection effect becomes more and more critical as we go towards the limb of the solar disk.
Therefore, in the first step, we have restricted ourselves up to $0.9R_{\odot}$.
Later on, to avoid the uncertainty in the magnetic field measurement we have also excluded the BMRs which have absolute mean heliographic longitude greater than 50$^\circ$ from our analysis.

To identify BMRs, 
we have followed the method given in \citet{SK12}.
So we have first applied a threshold on magnetic field strength and then a moderate flux balance condition to avoid the false detection of unipolar spot or BMR with large flux difference (see \Fig{fig:cont}(c--d)). 
Unlike \citet{SK12}, we have applied a 2D Gaussian smoothing with FWHM of 3 pixels \citep{Hgnr99} to reduce the spatial noise, before calculating (I) heliographic coordinate, (II) magnetic flux and (III) maximum field density from detected BMRs. Since maximum magnetic field density mimics the maximum field strength, we call it as the maximum field strength $B_\mathrm{max}$. While calculating $B_\mathrm{max}$ for HMI data, we have multiplied it by a factor of 1.4 to bring two data sets on the same scale \citep{2012SoPh..279..295L}. Tilts of BMRs have been calculated with respect to solar E--W direction considering the spherical geometry of the Sun.


\begin{figure*}[!htbp]
    \centering
    \includegraphics[width=0.8\textwidth]{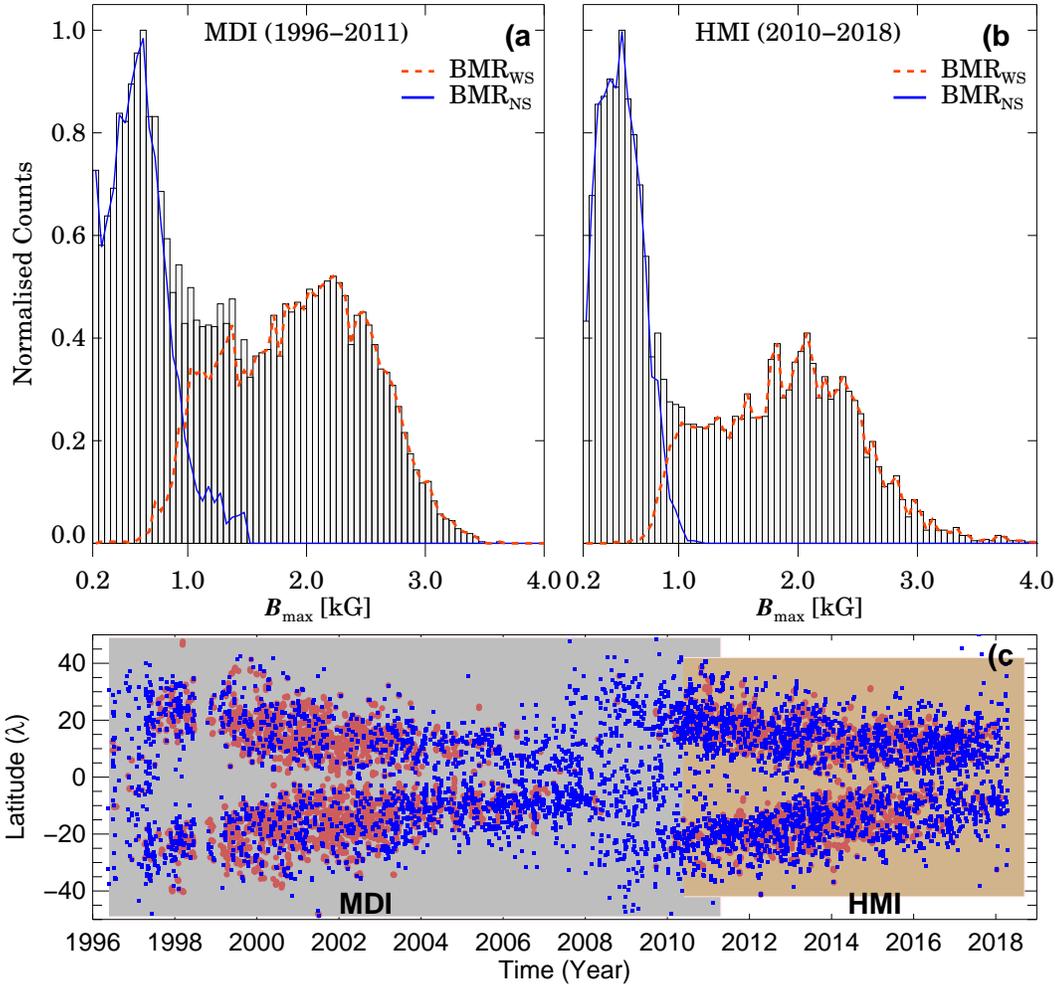}
    \caption{(a-b): Distributions of $B_{\rm max}$ in the BMRs from MDI (left panel) and HMI (right). 
             Red and blue respectively show $B_{\rm max}$ distributions of BMR$_{\rm WS}$ 
             (having counterpart in IC) and BMR$_{\rm NS}$ (no counterpart in IC).
             The vertical axes of two panes are divided by 315 and 388, respectively to bring the maxima of distributions to unity.
            Bottom: Time-latitude distribution of BMR$_{\rm WS}$ (red) and BMR$_{\rm NS}$ (blue).}
    \label{fig:BDist}
\end{figure*}

\section{Results and Discussions}
\label{sec:res}

Before we explore the magnetic field dependence of BMR tilt, we first present the 
distribution of the maximum magnetic field \Bmax\ of BMRs in \Fig{fig:BDist}(a-b). For the time being, we ignore the solid and dashed lines in these figures. We observe two well-separated peaks at around 600~G and 2100~G. These peaks are seen both in MDI and HMI data. HMI data includes the solar cycle 24, which is a relatively weak cycle and contains less number of strong field BMRs compare to weak field BMRs.
Despite the data obtained from two different instruments and two different solar cycles, we find the presence of two distinct peaks in both data sets.
These two distinct peaks remain even when we do not smooth the data or smooth with different windows.
However, as we smooth the data with a wider averaging window, these peaks tend to flatten out as well as shift slightly towards lower values. 
In the extreme limit, when we take the average magnetic field (i.e., window size equals to the BMR area), the two observed peaks 
disappear. This is expected because the magnetic field falls rapidly as we move away from the BMR center. 

 It appears that the whole HMI distribution is slightly shifted  to the left side and therefore the peaks appear at slightly smaller \Bmax\ than 
in MDI data. This could be due to different solar cycle, or it could be that the factor 1.4 
used to scale the HMI magnetic field is not appropriate for the entire range of \Bmax \citep{2012SoPh..279..295L, Pietarila13}.
Nevertheless, these results suggest that the magnetic field distribution of BMRs is bimodal 
and possibly there are two types of BMRs having significantly different field strength. 

To understand these two peaks in our data, we analyse their  IC for the same periods. The IC images may not necessarily be simultaneous but they are near-simultaneous with a maximum time difference being 3 hours.
We find that not all BMRs have their counterparts in IC (i.e., sunspots) (\Fig{fig:cont}(a) and \Fig{fig:cont}(f)). When we say counterpart in IC, we mean whether there is any spot present in the IC on the BMR region (as identified in the magnetogram), independent of their size.
It turns out that the BMRs which have their counterparts in IC (\Fig{fig:cont}(b) and \Fig{fig:cont}(f)) are having higher magnetic field.
When we overplot these two distributions in \Fig{fig:BDist}, we find that the BMRs having counterparts in IC (red/dashed line) beautifully represents the second peak at high $B_{\rm max}$ and the rest, i.e., BMRs without having a counterpart in IC (blue line), 
overlap with the first peak at the low $B_{\rm max}$.
Again we notice that in both the data sets this feature distinctly appears.
We define BMR$_{\rm WS}$ as the BMRs which have counterpart in IC, i.e., no sunspots and $B_{\rm max}$ distribution peaks at around 2~kG, while
BMR$_{\rm NS}$ as the BMRs which do not have sunspots (no counterpart in IC) and $B_{\rm max}$ distribution peaks at around 600~G.
Similar bi-modality in the maximum field distribution, have been reported in the past by \citet{cho2015} and \citet{Tla19} using sunspots and pores from SDO/HMI data. However in this work we look into the more general features, BMRs, of which sunspots and pores are part of.


\begin{figure}[!htbp]
    \centering
    \includegraphics[width=0.505\textwidth]{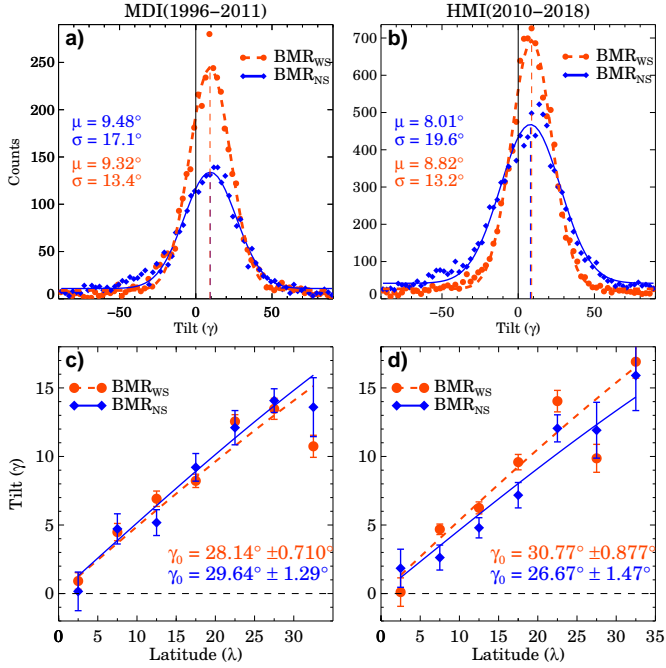}
    \caption{(a-b): Red and blue show tilt distributions of BMR$_{\rm WS}$ and BMR$_{\rm NS}$, respectively.
             Points represent the data and lines show the fitted Gaussians with parameters
             printed on the panels.
            (c-d): Mean tilt in each latitude bin as a function of the latitude. Solid and dashed lines are Joy's law ($\gamma = \gamma_0 \sin\lambda$) fits for BMR$_{\rm WS}$ and BMR$_{\rm NS}$.
            }
    \label{fig:tiltdist}
\end{figure}

Seeing the peak of BMR$_{\rm NS}$ at smaller field strength, one may conjecture that these
BMRs are produced from the small-scale magnetic field possibly originating  from the small-scale
dynamo \citep{PS93}. If this is the case, then we expect no preferred latitude distribution and no solar 
cycle variation. 
However, in \Fig{fig:BDist}(c), we find no such evidence. Both classes of BMRs
follow similar temporal and latitudinal variations in the usual butterfly diagram. Thus, this 
result do not suggest that the origin of BMR$_{\rm NS}$ are linked to the small-scale dynamo.

Now we explore the magnetic field dependence of BMR tilt. 
As we have found two distributions of BMRs, we shall first 
present the basic features of tilt of these two BMR classes separately. 
\Fig{fig:tiltdist} shows the tilt distributions of these two classes of BMRs namely, BMR$_{\rm WS}$ (red) and BMR$_{\rm NS}$ (blue) in the latitude range $10^\circ$--$~30^\circ$ including both the hemisphere.
Distributions peak at non-zero tilt and show Gaussian-like behaviour, which is of course not new \citep{WS89,SK12}. 
Although both distributions peak almost at the same tilt value, the distribution spreads are not identical 
and they are consistently different in two data sets. 
After fitting histograms with Gaussian profiles with mean $\mu$ and standard deviation $\sigma$, we find 
$\mu$ is around $9^{\circ}$ for both classes of BMRs and from both data sets. 
However, $\sigma$ for BMR$_{\rm WS}$ is smaller by a few degrees in both the data sets.
These results indicate that the tilt has some magnetic field dependence.

As shown in \Fig{fig:tiltdist}(c-d), Joy's law slope $\gamma_0$ are 
consistently different in two classes of BMRs. 
BMR$_{\rm WS}$ has a slightly larger $\gamma_0$ in HMI data,
while in MDI data it is opposite.
As MDI and HMI include data from two different times, 
we do not expect Joy's law trend to be identical in two data sets. 
Nonetheless, evidence of Joy's law in  BMR$_{\rm NS}$ further suggests that the BMR$_{\rm NS}$ class may not be originating
from the small-scale magnetic field, rather they must be originating from the same large-scale 
magnetic field which produces BMR$_{\rm WS}$.

\begin{figure*}[!htbp]
    \centering
    \includegraphics[width=1.8\columnwidth]{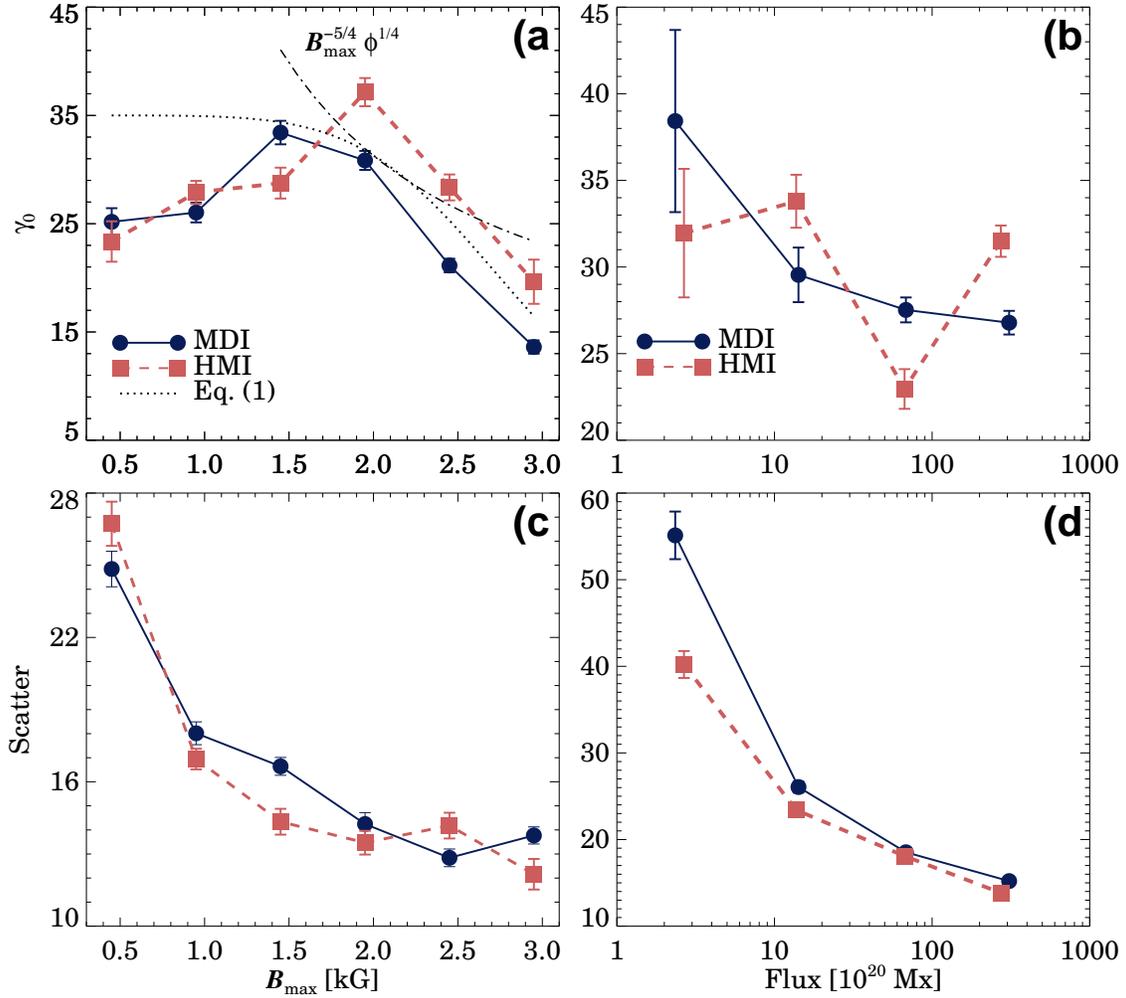}
    \caption{Magnetic field (\Bmax) dependences of: (a) Joy's law slope $\gamma_0$ 
              and (c) the tilt scatter $\sigma$. 
(b) and (d) are the same as left panels but as functions of flux.
}
    \label{fig:qu}
\end{figure*}

\subsection{Magnetic quenching of tilt angle}
\label{quenching}
To quantify the magnetic field dependence of BMR tilt, we now compute 
Joy's law slope $\gamma_0$ and the scatter around the mean tilt ($\sigma$), separately in each $B_{\rm max}$ bin with bin size of 500~G. 
In \Fig{fig:qu}(a), we observe that for MDI data, 
$\gamma_0$ is only slightly increased 
in the small \Bmax\ range and then dropped at least by about $15^\circ$ 
in the high field values above 2~kG.
While HMI data follow a general trend, there is a significant increase
in the low-field range.
A prominent reduction of $\gamma_0$ (by about $15^\circ$) with the magnetic field strength clearly establishes the existence of BMR tilt quenching. 
We emphasize that the tilt quenching is seen when \Bmax $> 2$~kG. That is 
why in \Fig{fig:tiltdist}, the mean Joy's law trend of BMR$_{\rm WS}$ is not smaller than BMR$_{\rm NS}$. It is only the strong BMRs$_{\rm WS}$ having \Bmax $> 2$~kG 
show the quenching in tilt.
 
We note that although the general trend of tilt quenching is seen, the results are slightly sensitive to the analysis, particularly, to the number of data.
We have checked that our results do not change when (i) taking different \Bmax\ bin, (ii) excluding data point if Joy's law fit
is not significant, and (iii) removing the data in the Joy's fitting if the BMR number is less than 50 in each latitude bin.
Further, the different beahviour of MDI and HMI always persists.
As seen in \Fig{fig:qu}(b), the variation with the BMR flux is monotonous for MDI data 
but not for HMI.

The indication of tilt quenching as seen in \Fig{fig:qu}(a) gives an observation
support of the following nonlinear quenching in the \bl\ $\alpha$
or in $\gamma_0$ routinely used to saturate the magnetic field growth in kinematic dynamo models \citep[e.g.,][]{CSD95, DC99, CNC04, KTV19}.
\begin{equation}
f_q \propto \frac{1}{\left[1 + (\frac{B_{\rm max} }{ B_0})^n \right]} 
\end{equation}
with $n=2$ \citep[see for example, Equation 10 of][]{KM17}.
However, our data fits best when $n = 5.8 \pm 0.8$ (and $B_0 = 2.9 \pm 0.1$~kG with reduced-$\chi^2 = 30.9$).

Now we discuss whether our results can be connected to the theory of thin flux tube model for the BMR formation.
Based on this theory, we expect, the intense toroidal flux rises fast, and thus, the Coriolis force gets less time to induce a tilt. 
Hence, the BMR tilt is expected to decrease with the increase of the magnetic field. The thin flux tube simulations of \citet{FFM94} predicted:
\begin{equation}
\gamma \propto \sin\lambda B_0^{-5/4}\Phi^{1/4},
\label{eq1}
\end{equation}
where $B_0$ is the initial magnetic field of the toroidal flux tube and
$\Phi$ is the flux content.
The theoretical study suggests that due to combined effects of rapid expansion, radiative cooling, and pressure buildup,
the magnetic fields of BMRs forming loops become sufficiently low as they rise towards the surface,
and within a few Mm depth BMRs tends to get disconnected from their roots
\citep{SR05}.
The current understanding of the whole process is 
very limited; however see \citet{RC14,FF14,Nel14}.
Therefore, we do not know whether the initial magnetic field $B_0$ is related to 
the \Bmax\ that
we observe inside the BMR. 
However, if we assume that $B_0 \propto$ \Bmax, then we can make some comment on the thin flux tube model. 

The BMR flux $\Phi$ is observed to vary with the
magnetic field strength \citep{Tla14}. 
In our data, we find the following relation hold resonably well.
\begin{equation}
\frac{\Phi} {\angles{\Phi}} = a + b \frac{B_{\rm max}} {\angles{B_{\rm max}}} + c \left( \frac{B_{\rm max}} {\angles{B_{\rm max}}} \right) ^2+ d \left( \frac{B_{\rm max}} {\angles{B_{\rm max}}} \right) ^3,
\label{eq:PhiB}
\end{equation}
where $a=-0.08 \pm 0.01$, $b=0.84 \pm 0.12$,  $c=-0.57 \pm 0.19$ and $d= 0.52 \pm 0.08$ for MDI data and $a=-0.09\pm 0.02$, $b=0.81 \pm 0.15$,
$c=-0.27 \pm 0.23$ and $d= 0.32 \pm 0.10$ for HMI. 
Putting this relation in \Eq{eq1}, we find that the slope of Joy's law $\gamma_0$ decreases as shown by the dashed line.
We observe that in the high-field regime, our result qualilatively supports the thin flux tube model.

In the low-field regime with $B_{\rm max} < 2$~kG 
$\gamma_0$ increases with \Bmax\ which does not fit with the thin flux tube model. However,
we should not forget that this model does not include the convection, which can affect the dynamics of the flux tube to
change the tilt through the helical convection.
By considering convection, in the thin flux tube model,
\citet{WFM11} showed that while the general Joy's law trend is recovered,
the tilt increases with the increase of magnetic field strength first in the low
field regime, and then it reduces; see their Figure~8 and 12 \citep[also see][]{WFM13}. 
Similar behaviour is found in our data; see \Fig{fig:qu}(a).

Thin-flux tube rise model also predicted that the rising-flux loops could be buffeted by the turbulent convection
during their rise in the CZ and this could cause a scatter
around the systematic tilt variations --- Joy's law \citep{LF96, LC02}.
When the magnetic field is strong, we expect the magnetic tension to oppose this buffeting of flux tubes and
the scatter to be less. Further, strong flux tubes rise faster (due to strong magnetic buoyancy)
and thus they get less time to be buffeted by  convection \citep{WFM11}.
The tilt scatter computed from our data supports this idea. In \Fig{fig:qu}(c-d),
we see that it systematically decreases with the increase of \Bmax\ or flux. 

\section{Conclusion}
In this Letter, we have studied BMRs detected from the magnetograms of
MDI (1996--2011) and HMI (2010--2018). In both the data sets, we find that the BMR number distribution
shows a bimodal distribution when measured with respect to their maximum magnetic field \Bmax.
The first peak at low field (\Bmax$\approx600$~G) corresponds to BMRs which do not have
counterparts in IC (i.e., no sunspots),
while the second peak at high field (\Bmax$\approx 2100$~G) corresponds to BMRs
which have counterparts in IC. 
BMR$_{\rm NS}$ also shows a similar butterfly diagram, tilt distribution and Joy's law
as that of BMR$_{\rm WS}$. This suggests that BMR$_{\rm NS}$ are not produced from the small-scale magnetic field, rather they must be produced from the same large-scale global field which produces sunspots.
One difference between these two classes of BMRs is that the tilt scatter and
the slope of Joy's law $\gamma_0$ are smaller in BMR$_{\rm WS}$. 
However, our study does not explain why BMR show two distinct peaks in the $B_{\rm max}$ distribution, which requires further studies.

On computing the tilt in each \Bmax\ bin, we find a significant change in the BMR tilt for MDI and HMI data.
In the low \Bmax\ range, 
$\gamma_0$ increases with the increase of \Bmax.
However, for $B_{\rm max} > 2$~kG (which corresponds to strong sunspots), 
$\gamma_0$ decreases with \Bmax.
These results are in 
qualitative
agreement with the predictions of the thin flux-tube rise 
model \citep{DC93,FFM94,CMS95,Fa09} and in particular the simulations 
with the convection \citep{WFM11,WFM13}. The reduction of tilt with the increase of the magnetic field 
in the high field regime gives a hint for the nonlinear quenching routinely used in the \bl\ type kinematic dynamo models. 

We understand that the variations of BMR properties, particularly the tilt quenching with magnetic field are demonstrated in a relatively narrow range. This, however, is due to the fact that the availability of data are limited and Joy's law is a statistical relation. Furthermore, the last two cycles, during which our analyses are performed, are relatively weak, having weak BMR field strength. The highest magnetic field in our BMRs data is about 3~kG, and the magnetic quenching is expected to be more in the super-kilogauss magnetic field. Therefore, we believe that our results need to be investigated further with larger data sets, especially from stronger cycles having high-field BMRs.

\acknowledgements
We thank the anonymous referee for the detailed comments which helped to clarify some issues in the revised manuscript.
 BBK sincerely acknowledges financial support from Department of Science and Technology
(SERB/DST), India through the Ramanujan Fellowship (project no SB/S2/RJN-017/2018).
He further appreciates loving hospitality at Indian Institute of Astrophysics, Bangalore during this project. 

\bibliographystyle{apj}


\end{document}